\documentstyle[aps,epsfig,multicol,times]{revtex}

\begin{document}

\draft

\title{\protect\vspace*{15mm}\LARGE\bf
Phase Shifts of Atomic de Broglie Waves\\
at an Evanescent Wave Mirror}
\author{%
C. Henkel\thanks{\protect
Presently at Institut f\"{u}r Physik,
Universit\"{a}t Potsdam, 14469 Potsdam, Germany,
email Carsten.Henkel@quantum.physik.uni-potsdam.de
\hspace*{05mm}\hskip\fill
{\it Retyping courtesy of M. Path and R. Donner\/}}%
, J.-Y. Courtois, R. Kaiser, C. Westbrook, and
A. Aspect}
\address{Institut d'Optique Th\'eorique et Appliqu\'ee,
B.P. 147, F-91403 Orsay, France}
\date{\protect Received October 21, 1993}
\maketitle

\begin{abstract}
\hspace*{-3ex}{\bf Abstract --} 
A detailed theoretical investigation of the reflection of an atomic 
de Broglie wave at an 
evanescent wave mirror is presented. The classical and the 
semiclassical descriptions of the 
reflection process are reviewed, and a full wave-mechanical approach 
based on the analytical 
soution of the corresponding Schr\"odinger equation is presented. The 
phase shift at reflection is 
calculated exactly and interpreted in terms of instantaneous 
reflection of the atom at an 
effective mirror. Besides the semiclassical regime of reflection 
describable by the WKB method, 
a pure quantum regime of reflection is identified in the limit where 
the incident de Broglie 
wavelength is large compared to the evanescent wave decay length.
\end{abstract}


\vspace*{-100mm}
\begin{footnotesize}\it\noindent
Laser Physics, Vol. 4, No. 5, 1994, pp. 1042 - 1049
\\[-0.9mm]
Original Text Copyright \copyright 1994 by Astro Ltd.
\\[-0.9mm]
Copyright \copyright 1994 by MAHK Hayka/Interperiodica Publishing 
(Russia).
\end{footnotesize}

\setcounter{page}{1042}

\vspace*{-8mm}
\begin{flushright}
\rule[0.8ex]{63mm}{0.9pt}%
\hspace*{-63mm}%
\rule{63mm}{0.9pt}%
\hspace*{1ex}
\large\bf ATOM OPTICS
\end{flushright}

\begin{center}\large\bf
SPECIAL ISSUE
\\
``LASER COOLING AND TRAPPING''
\end{center}

\vspace*{65mm}

\begin{multicols}{2}
\narrowtext

\section{Introduction}

\vspace*{-02mm}

Atomic mirrors are one of the key components in the field of atom 
optics [1]. In order to realize 
such a device, the use of evanescent waves appears very promising [2, 
3]. In view of the future 
applications fo evanescent wave mirrors, a detailed investigation of 
their optical properties 
seems appropriate. For some basic purposes (for example, the 
deflection of an atomic beam), the 
characterization of the geometric optical properties of the 
evanescent wave mirror is sufficient. 
This can be achieved by treating the incident atom as classical 
particles and by deriving their 
classical trajectories (this is analogous to the calculation of light 
rays in conventional 
optics). For more elaborate purposes (atom interferometers [4], 
atomic cavities [5, 6]), knowledge 
of the wave-mechanical properties of the evanescent wave mirror is 
also required. One then needs 
to describe the atom by a de Broglie wave in order to estimate the 
phase shift experienced by the 
atom during reflection at the mirror. A semiclassical derivation of 
this phase shift, based on the 
evaluation of the action integral along the classical atomic 
trajctories (WKB method), has been 
given by Opat {\it et al.\/} [7]. We present in this paper a more 
complete approach 
based on the analytical solution of the Schr\"odinger equation 
describing the interaction between 
the atom and the evanescent wave mirror in the regime of coherent 
atom optics (limit of negligible 
spontaneous emission). We interpret the atomic phase shift derived 
from the atomic wave function 
in terms of instantaneous reflection at an effective mirror, which 
generalizes the one introduced 
in [7]. We distinguish between a semiclassical and a quantum regime 
of reflection. In the 
semiclassical regime, realized at high incident energy, the 
Schr\"odinger and the WKB approach 
coincide, and the evanescent wave mirror behaves as a dephasing 
dispersive mirror, analogous to a 
dielectric mirror in conventional optics. By contrast, in the quantum 
regime of reflection where 
the incident atomic de Broglie wavelength is large compared to the 
evanescent potential decay 
length, the evanescent wave mirror acts as a nondispersive infinitely 
steep barrier, analogous to 
a metallic mirror in conventional optics. Finally, the reflection 
process of an atomic wave packet 
incident on an evanescent wave mirror is discussed.

\section{Classical and semiclassical descriptions of the reflection 
process}
\vspace*{-02mm}

Before turning to the full wave-mechanical treatment of atomic 
reflection at an evanescent wave 
mirror, we start by reviewing the classical dynamics as well as the 
WKB description of the 
reflection process. This will allow us to make a clear distinction 
between the semiclassical and 
the pure quantum features of atomic reflection.

\subsection{Presentation of the Model}
\vspace*{-02mm}

We consider the simple case of a two-level atom normally incident of 
the surface ($z=0$ plane) of 
an evanescent wave mirror.\footnote{Because of the translational 
symmetry in the directions 
parallel to the mirror surface, the problem can be reduced to one 
dimension. This simplification 
holds for both the classical and the quantum viewpoints.} Because we 
are interested in the regime 
of coherent atom optics (limit of negligible spontaneous emission), 
we restrict ourselves to the 
limit of low saturation of the atomic transition, where the reactive 
part of the atom-evanescent 
wave coupling (light shift) is predominant over the dissipative part. 
We also assume that the 
detuning between the evanescent wave and the atomic frequency is 
properly chosen so that the atom 
ca be considered to follow adiabatically the optical potential 
associated with the light-shifted 
ground-state level. In this regime, all the physical phenomena can be 
accounted for by means of 
the Hamiltonian [3, 7]:
\begin{equation}
H = \frac{p^2}{2M} + \frac{p_{\rm max}^2}{2M} \exp (-2 \kappa z),
\end{equation}
\noindent
which contains the atomic kinetic energy (first term) and the 
reactive part of the atom-field 
coupling (second term). In equation (1), $p$ and $z \ge 0$ are the 
momentum and position of the 
atomic center of mass, $M$ is the atomic mass, $p_{\rm max} > 0$ is 
the maximum momentum that can be 
reflected by the optical potential barrier, and $\kappa^{-1}$ is the 
characteristic decay length 
of the evanescent wave, of the order of the laser wavelength%
\footnote{Note that, this estimate 
breaks down near the critical angle for the evanescent laser wave. In 
this case, the length scale 
$\kappa^{-1}$ tends to infinity.} (for a discussion of typical 
experimental parameters, see 
Appendix A). Note that when quantizing the atomic external degrees of 
freedom, one has to 
substitute the momentum and position operators $P$ and $Z$ for $p$ 
and $z$ in equation (1). The 
potential in equation (1) grows exponentially as $z \to - \infty$. We 
thus neglect any effects due 
to tunneling through the potential barrier to the physical mirror 
surfce.

\subsection{Classical Dynamics of the Reflection Process}
\vspace*{-02mm}

Let us first consider the incident atom as a classical particle with 
asymptotic momentum 
$-p_{\infty} (0 < p_{\infty} < p_{\rm max})$. With an appropriate 
choice of time origin, the 
classical trajectory of the atom can be written [7] (Fig. 1):
\begin{equation}
z(t) = z_0 + \kappa^{-1} \ln \cosh (t/\tau_{\rm refl}),
\end{equation}
\noindent
where
\begin{equation}
z_0 = \kappa^{-1} \ln (p_{\rm max} / p_{\infty})
\end{equation}
\noindent
is the position of the turning point of the turning point of the 
trajectory (reached at $t=0$), 
located about $\kappa^{-1}$ in front of the mirror surface, and where
\begin{equation}
\tau_{\rm refl} = M / \kappa p_{\infty}
\end{equation}
\noindent
is the time scale for the reflection process and corresponds to the 
time taken to cross the 
thickness $\kappa^{-1}$ of the optical potential at the asymptotic 
velocity $p_{\infty} / M$ (see 
Appendix A, the table for typical experimental values).

In the asymptotic region $z \gg \kappa^{-1}$ of vanishing optical 
potential, the atom is moving 
freely at constant velocity $\mp p_{\infty} / M$ along the asymptotes 
of the classical trajectory 
(see Fig. 1). These straight asymptotes intersect at the position 

\begin{figure}[tbh]
\centerline{\epsfig{file=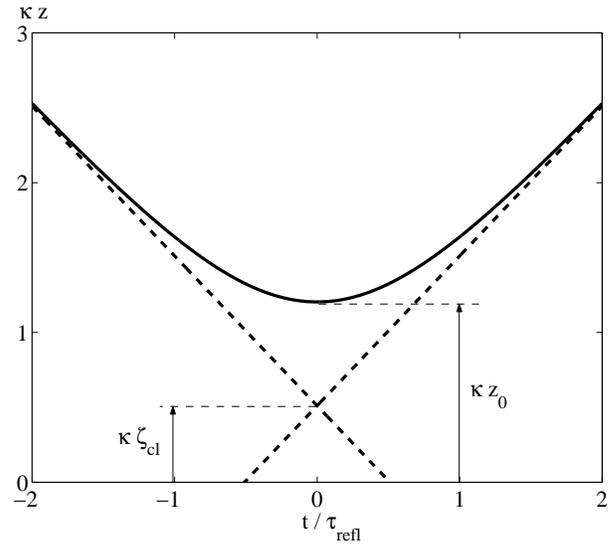,width=80mm}}
\caption[]{%
\noindent
Classical trajectory of an atom undergoing specular reglection at an 
evanescent wave 
mirror. The dimensionless atomic position $\kappa z$ is represented 
vs.\ $t/\tau_{\rm refl}$ for the 
parameters $p_{\infty} = 3 \hbar \kappa$ and $p_{\rm max} = 10 \hbar 
\kappa$. The atom approaches the 
mirror surface at the minimum distance $z_0 \approx 1.2 \kappa^{-1}$ 
and is reflected on a time 
scale of the order of $\tau_{\rm refl}$. In the asymptotic region 
($\kappa z \gg 1$) of vanishing 
optical potential, the atom propagates freely at the velocity $\pm 
p_{\infty} / M$, and the 
classical trajectory corresponds to straight asymptotes (dashed 
lines), which intersect at the 
position $\zeta_{\rm cl} \approx 0.5 \kappa^{-1}$.
}
\end{figure}
\begin{equation}
\zeta_{\rm cl} (p_{\infty}) = z_0 (p_{\infty}) - \kappa^{-1} \ln 2
,
\end{equation}
\noindent
which is shifted more deeply into the potential relative to $z_0$ by 
a quantity independent of the 
incident momentum (see Fig. 1). As far as the asymptotic classical 
dynamics of the atom is 
concerned, the evanescent wave mirror thus behaves as a fictitious 
infinitely steep barrier 
located at $\zeta_{\rm cl} (p_{\infty})$, at which the atom would 
experience an instantaneous 
reflection. We name this barrier the classical effective mirror after 
[7].

\subsection{The WKB Solution for the Evanescent Wave Mirror}
\vspace*{-02mm}

Let us now consider the semiclassical description of the reflection 
process. In this case, the 
atom is described by means of a wave function derived using the WKB 
approximation. In the 
classical allowed region $z > z_0$, this WKB wave function is given 
by [8]:
\begin{eqnarray}
&& \psi_{\rm WKB}(z) = \sqrt{\frac{4M}{p(z)}} \sin \left( \frac{\pi}{4} + 
\frac{1}{\hbar}\int_{z_0}^z 
p(z') dz' \right) 
\\
&& = \sqrt{\frac{4M}{p(z)}} \sin \left[ \frac{\pi}{4} + 
\frac{p_{\infty}}{\hbar \kappa}(\mbox{artanh}
 (p(z)/p_{\infty}) - p(z)/p_{\infty}) \right], 
\nonumber
\end{eqnarray}
\noindent
where $p(z)$ is the classical momentum calculated from energy 
conservation:
\begin{equation}
p(z)^2 + p_{\rm max}^2 \exp(-2 \kappa z) = p_{\infty}^2,
\end{equation}
\noindent
and where the phase $\pi/4$ results from the WKB connection formula, 
which matches the 
oscillating part of the wave function (6) to the decaying part in the 
classically forbidden 
region $z < z_0$. The normalization of the WKB wave function (6) has 
been chosen such that the 
incident and reflected waves both have unit flux independent of the 
asymptotic momentum.

In the asymptotic region, the atomic wave function is a superposition 
of two plane waves with wave 
vectors $k_{\infty} = \mp p_{\infty}/\hbar$, which correspond to the 
incident and reflected waves. 
The phase shift that we are interested in at reflection $\Delta 
\varphi_{\rm WKB}$ is related to the 
relative phase between these two plane waves. We define $\Delta 
\varphi_{\rm WKB}$ by writing the 
asymptotic WKB wave function in the form:
\begin{equation}
z \to + \infty : \psi_{\rm WKB}(z) \cong \sqrt{\frac{4M}{p_{\infty}}} 
\sin \left( \frac{1}{\hbar} 
p_{\infty} z + \frac{1}{2} \Delta \varphi_{\rm WKB} \right).
\end{equation}
\noindent
This definition of the phase shift takes as reference ($\Delta 
\varphi_{\rm WKB} = 0$) a standing wave 
in front of an infinitely steep barrier located at the origin $z = 
0$. It is important to note 
that this definition is somewhat arbitrary. In [7], for example, the 
reference is a standing wave 
in fromt of an infinitely steep barrier located at the position $z = 
\zeta_{\rm cl}$ of the effective 
classical mirror [equation (5)]. Because $\zeta_{\rm cl}$ depends on 
the incident atomic momentum, 
however, this phase reference is not absolute. With our definition of 
the phase shift, $\Delta 
\varphi_{\rm WKB}$ is the phase correction in the situation where the 
evanescent optical potential is 
approximated by an infinitely steep barrier located at the mirror 
surface [6]. By writing the 
asymptotic expansion of the WKB wave function (6) in the form (8), 
one obtains [7]
\begin{eqnarray}
\Delta \varphi_{\rm WKB}(p_{\infty}) & = & \frac{\pi}{2} - 
2\frac{p_{\infty}}{\hbar \kappa} \left[ 1 + \ln 
\left( \frac{p_{\rm max}}{2p_{\infty}} \right) \right]
\\
& = & \delta 
\varphi_{\rm WKB} - 2 p_{\infty} \zeta_{\rm WKB} / \hbar 
\nonumber
\end{eqnarray}
\noindent
with
\begin{mathletters}
    \begin{eqnarray}
\delta \varphi_{\rm WKB} & = & \pi / 2 
\\ 
\zeta_{\rm WKB}(p_{\infty}) &=& \zeta_{\rm cl}(p_{\infty}) + \kappa^{-1}. 
\end{eqnarray}
\noindent
The order of magnitude of the WKB phase shift (9) is given by the 
ratio $p_{\infty} / \hbar 
\kappa$, which represents the phase shift associated with the free 
propagation of an atom of 
momentum $p_{\infty}$, through the spatial extent $\kappa^{-1}$ of 
the evanescent optical 
potential. 

The physical interpretation of (10) becomes transparent if we write 
the asymptotic WKB wave 
function (8) as%
\footnote{This decomposition separates the phase 
shift into a constant and an 
essentially linear term. Note that such an interpretation is not 
always unambiguous because one 
has to decompose $\Delta \varphi(p_{\infty}) = \delta \varphi - 2 
p_{\infty} \zeta / \hbar$, where 
$\delta \varphi \in [0, 2\pi]$ and $\zeta$ are weakly dependent on 
$p_{\infty}$.} 
\end{mathletters}
\begin{eqnarray}
&& z \to + \infty: 
\nonumber\\
\psi_{\rm WKB}(z) &\cong& \sqrt{\frac{4M}{p_{\infty}}} 
\sin \left( \frac{1}{2} \delta 
\varphi_{\rm WKB} + \frac{1}{\hbar} p_{\infty}
(z - \zeta_{\rm WKB}) \right),
\nonumber\\
&&
\end{eqnarray}
\noindent
which corresponds to a plane standing wave whose phase is fixed to 
the value $\frac{1}{2} \delta 
\varphi_{\rm WKB}$ at $z = \zeta_{\rm WKB}$. As far as the asymptotic WKB 
wave function is concerned, the 
evanescent wave mirror is thus equivalent to an effective dephasing 
mirror located at the position 
$\zeta_{\rm WKB}$, where the atomic wave is instantaneously reflected and 
phase shifted by the amount 
$\delta \varphi_{\rm WKB}$ (as light on a mirror). By analogy with the 
classical case (Section 2.2), 
we call this mirror the WKB effective mirror. The dephasing character 
of this mirror results from 
the WKB phase factor $\pi/4$ and thus has a nonclassical origin. It 
holds as long as the WKB 
approximation remains valid, i.e., as long as the incident de Broglie 
wavelength is small compared 
to the decay length $\kappa^{-1}$ of the evanescent optical potential 
[8]. Furthermore, the 
evanescent wave mirror is dispersive because of the dependence of 
$\zeta_{\rm WKB}$ on $p_{\infty}$. 
More precisely, the condition for the mirror to be dispersive is that 
$\partial^2 \Delta \varphi_
{WKB} / \partial p_{\infty}^2 \neq 0$: a linear dependence of $\Delta 
\varphi_{\rm WKB}$ on 
$p_{\infty}$ can always be removed by an appropriate choice of 
absolute phase reference. Finally, 
it is interesting to note that the position 
$\zeta_{\rm WKB}$ [equation (10b)] of the WKB effective mirror
differs from the classical effective mirror position $\zeta_{\rm cl}$
[equation (5)] by the quantity 
$\kappa^{-1}$ independent of the incident atomic momentum. As a 
result, the classical and the WKB 
description of the reflection process yield comparable physical 
pictures.

\section{Schr\"odinger wave function approach}
\vspace*{-02mm}

We now turn to the full wave-mechanical description of atomic 
reflection at the evanescent wave 
mirror. This description is based on the analytical solution of the 
corresponding Schr\"odinger 
equation, which allows an exact calculation of the phase shift at 
reflection. 

\subsection{Solution of the Stationary Schr\"odinger Equation}
\vspace*{-02mm}

The full quantum description of atomic reflection consists in solving 
exactly the stationary 
Schr\"odinger equation for the atomic wave function $\psi(z)$:
\begin{equation}
\left( -\hbar^2 \frac{d^2}{dz^2} + p_{\rm max}^2 \exp (-2 \kappa z) 
\right) \psi (z) = p_{\infty}^2 
\psi (z).
\end{equation}
\noindent
We use the change of variable

\begin{equation}
z \to u = \frac{p_{\rm max}}{\hbar \kappa} \exp (- \kappa z)
\end{equation}
\noindent
which takes advantage of the invariance of the Hamiltonian (1) under 
the transformation

\begin{equation}
\forall a, \qquad \left\{ \begin{array}{lcl}  
z \to z + a \\ p_{\rm max} \to e^{2 \kappa a} p_{\rm max} 
\end{array} \right. .
\end{equation}

Equation (12) transforms into a Bessel-type equation:

\begin{equation}
\left( u^2 \frac{d^2}{du^2} + u \frac{d}{du} - (u^2 - \alpha^2) 
\right) \psi (u) = 0,
\end{equation}
\noindent
which only depends on one dimensionless parameter:

\begin{equation}
\alpha = p_{\infty} / \hbar \kappa.
\end{equation}

The solutions of (15) are linear combinations of the Bessel functions 
$I_{\pm i \alpha} (u)$. Two 
boundary conditions impose a unique solution:

(i) The wave function must vanish in the limit $z \to - \infty$ (the 
probability of the atom¥s 
being in the region $z \le z_0$ inside the potential being small 
compared to the probability of 
being in the classically allowed region $z \ge z_0$).

(ii) In the asymptotic region $z \to + \infty$, the wave function is 
normalized in the same way 
as the WKB solution [equation (6)].

As shown in Appendix B, these conditions lead to the solution%
\footnote{We have neglected any loss 
resulting from atomic tunneling to the mirror surface. However, the 
tunneling probability can be 
estimated by the flux of the wave function at $z=0$.
{\it Note added after publication:} To our knowledge,
the exact solution~(17)
for the exponential barrier has been first derived
by J. M. Jackson and N. F. Mott in 1932, {\it Proc.\ Roy.\ Soc.\
(London) Ser.\ A\/} {\bf 137}, 703.}
\begin{equation}
\psi_{\rm Schr} (z) = \sqrt{ \frac{4M}{p_{\infty}} 
\frac{\pi\alpha}{\sinh(\pi\alpha)}} \frac{1}{2i} 
(I_{-i\alpha} (u(z)) - I_{i\alpha} (u(z))).
\end{equation}

The Schr\"odinger wave function (17) and the corresponding WKB wave 
function (6) are represented 
in Fig. 2 as a function of the dimensionless parameter $\kappa z$, in 
the case $p_{\infty} = 3 
\hbar \kappa$ and $p_{\rm max} = 10 \hbar \kappa$. One sees that the 
wave functions are in good 
agreement in both the asymptotic region ($\kappa z \gg 1$) and far 
inside the optical potential 
($\kappa z \ll 1$), but that they significantly differ around the 
classical turning point 
$\kappa z_0 \cong 1.2$ (where the WKB wave function actually 
diverges).

\begin{figure}[bth]
\centerline{\epsfig{file=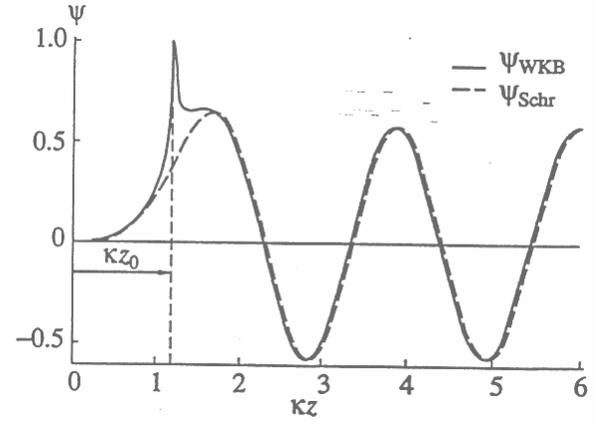,width=80mm}}
    \caption[]{%
Comparison between the WKB ($\psi_{\rm WKB}$) and the 
Schr\"odinger ($\psi_{\rm Schr}$) wave 
functions for the same parameters as in Fig.1. These wave functions 
coincide both in the asyptotic 
region ($\kappa z \gg 1$) and far inside the optical potential 
($\kappa z \ll 1$) but 
significantly differ around the classical turning point $z_0$,
where $\psi_{\rm WKB}$ diverges.
}
\end{figure}

\subsection{Phase Shift of the Schr\"odinger Wave Function }
\vspace*{-02mm}

The exact solution (17) of the Schr\"odinger equation allows us to 
derive exactly the phase shift 
experienced by the atomic wave function at reflection on the 
evanescent wave mirror.

Following the definition (8) of the WKB phase shift at reflection, we 
define the Schr\"odinger 
phase shift $\Delta \varphi_{\rm Schr}$ by writing the asymptotic 
Schr\"odinger wave function (17) in 
the form:
\begin{equation}
z \to + \infty: \psi_{\rm Schr} (z) \cong \sqrt{\frac{4M}{p_{\infty}}} 
\sin \left( \frac{1}{\hbar} 
p_{\infty} z + \frac{1}{2} \Delta \varphi_{\rm Schr} \right).
\end{equation}
\noindent
By using (17) and the asymptotic expansion ($u \to 0$) of the Bessel 
functions $I_{\pm i \alpha} (u)$, 
one obtains (see Appendix B)
\begin{equation}
\Delta \varphi_{\rm Schr} (p_{\infty}) = - 2 \alpha \ln \left( 
\frac{p_{\rm max}}{2 \hbar \kappa} \right) 
+ 2 \arg \Gamma (1+i\alpha),
\end{equation}
\noindent
where $\Gamma$ is the Euler gamma (factorial) function [9], and where 
$\arg\Gamma(1+i\alpha)$ is 
the argument of the complex number $\Gamma(1+i\alpha)$ defined as a 
continuous function of 
$\alpha$.

The exact ($\Delta \varphi_{\rm Schr}$) and the semiclassical ($\Delta 
\varphi_{\rm WKB}$) phase shifts 
at reflection are represented in Fig. 3 as a function of the 
dimensionless parameter $\alpha = 
p_{\infty}/\hbar\kappa$. One can clearly distinguish between two 
limiting cases.

In the limit $\alpha \gg 1$ (high incident momentum), where the 
incident atomic de Broglie 
wavelength is small compared to the decay length $\kappa^{-1}$ of the 
optical potential, the WKB 
and the Schr\"odinger approaches yield comparable phase shifts [8]. 

\clearpage

\widetext

\begin{figure}[tbh]
\centerline{\epsfig{file=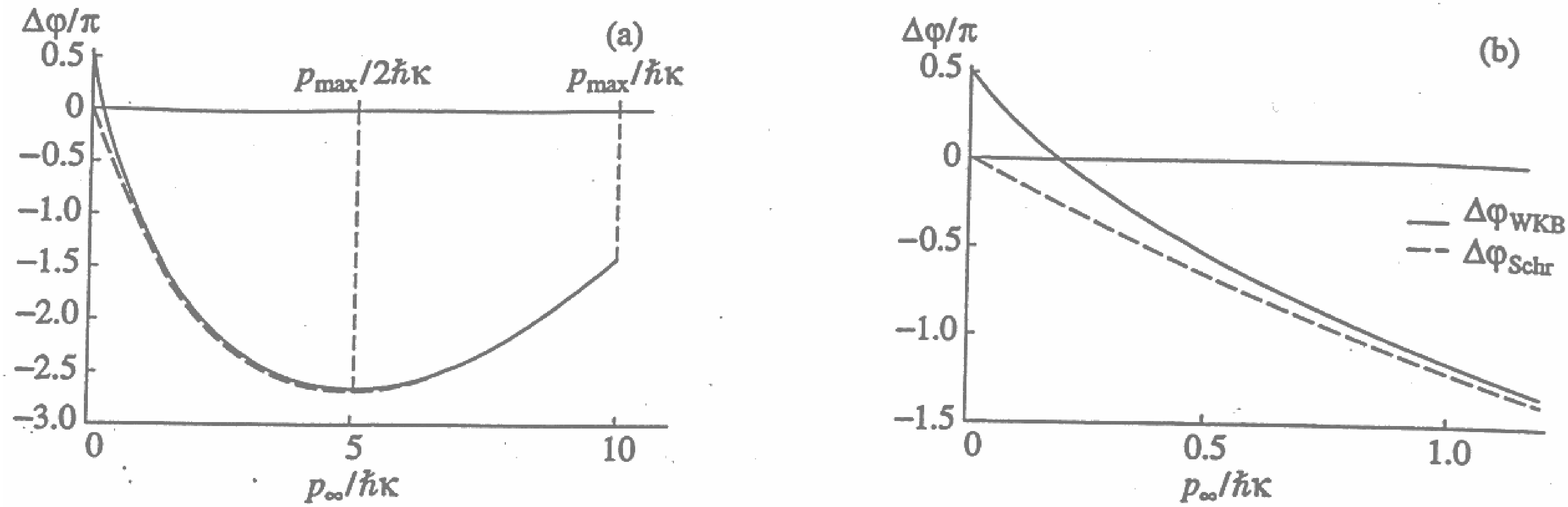,width=170mm}}
    \caption[]{%
Dependence of the WKB ($\Delta \varphi_{\rm WKB}$) and of the 
Schr\"odinger ($\Delta \varphi_{\rm Schr}$) 
phase shifts vs.\ the dimensionless incident atomic momentum 
$\alpha = p_{\infty}/\hbar
\kappa$ for $p_{\rm max} = 10\hbar\kappa$. (a) $\Delta \varphi_{\rm WKB}$ 
and $\Delta \varphi_{\rm Schr}$ 
coincide in the limit of high incident atomic momentum ($\alpha \gg 
1$), which thus corresponds to 
a semiclassical regime of reflection. (b) In the limit of low 
incident atomic momentum ($\alpha 
\ll 1$), the semiclassical description of the evanescent wave mirror 
does not yield the correct 
phase shift at reflection. This corresponds to a pure quantum regime 
of reflection.
}
\end{figure}

\narrowtext
\noindent
This corresponds to the 
semiclassical regime of atomic reflection considered in Section 2.3. 
Using equation (19), it is 
possible to derive the first correction to the semiclassical phase 
shift. One thus finds [9]
\begin{equation}
\alpha \gg 1: \Delta \varphi_{\rm Schr} (p_{\infty}) = \Delta 
\varphi_{\rm WKB} - \frac{1}{6\alpha} + 
O(\alpha^{-3}).
\end{equation}

In the limit $\alpha \ll 1$ (low incident momentum), where the 
incident atomic de Broglie 
wavelength is larger than the decay length of the evanescent optical 
potential, the WKB and the 
Schr\"odinger approaches yield different phase shifts. In particular, 
in the limit $\alpha \to 
0^+$, $\Delta \varphi_{\rm WKB}$ tends toward $\pi/2$, whereas $\Delta 
\varphi_{\rm Schr}$ tends toward 
$0$ (see Fig.\ 3b). The limit of low incident momentum thus 
corresponds to a pure quantum regime of 
reflection, which can not be appropriately described in semiclassical 
terms. In order to obtain in 
this regime a representation of the evanescent wave mirror in terms 
of an effective mirror, we 
write equation (19) in a form analogous to (9). One thus finds [9]
\begin{eqnarray}
&&\alpha \ll 1:
\nonumber\\
\Delta \varphi_{\rm Schr} &\cong& -2\alpha \left( \gamma + 
\ln \left( \frac{p_{\rm max}}{2 
\hbar \kappa} \right) \right) 
\\
&=& \delta \varphi_{\rm Schr} - 2 p_{\infty} 
\zeta_{\rm Schr} / \hbar
\nonumber
\end{eqnarray}
\noindent
with the Euler constant $\gamma \cong 0.577$ and
\begin{mathletters}
\begin{eqnarray}
\delta \varphi_{\rm Schr} &\cong& 0 
\\ 
\zeta_{\rm Schr}(p_{\infty} \ll \hbar \kappa) 
&\cong& \zeta_{\rm cl}(\hbar 
\kappa) + \gamma \kappa^{-1}.  
\end{eqnarray}

Equations (21) and (22) show that, in the quantum regime of 
reflection, the evanescent wave mirror 
behaves as a nondephasing effective mirror located at the position $z 
\approx \zeta_{\rm cl}(\hbar
\kappa)$

\parbox[t]{40mm}{%
\vspace*{74mm}}

\noindent
[equation (22b)], where the atomic wave is instantaneously 
reflected (as light on a 
metallic mirror). The nondephasing character of the Sch\"odinger 
effective mirror [equation (22a)] 
results from the fact that, on the spatial scale of the incident de 
Broglie wavelength, the 
evanescent optical potential appears to be an infinitely steep 
barrier. In the quantum regime of 
reflection, it is therefore legitimate to approximate the evanescent 
wave mirror by a hard 
barrier located at the position of the classical effective mirror for 
an asymptotic momentum 
$p_{\infty} \approx \hbar \kappa$. It is also interesting to note 
that, contrary to the 
semiclassical case, the position $\zeta_{\rm Schr}$ of the Schr\"odinger 
effective mirror is 
essentially independent of $p_{\infty}$. As a result, the evanescent 
wave mirror is no longer 
dispersive in the quantum regime.%
\footnote{It is perhaps surprising 
that the evanescent wave 
mirror is not despersive while the Schr\"odinger phase shift (21) 
depends linearly on the incident 
momentum. In fact, this dependence is related to the choice of 
reference for the phase shift. 
Thus, by taking as phase reference a standing wave in front of an 
infinitely steep barrier located 
in $z \cong \zeta_{\rm cl}(\hbar\kappa) + \gamma\kappa^{-1}$, the 
phase shift at reflection in the 
quantum regime would be independent of the incident momentum.}
\end{mathletters}

\section{Reflection of an atomic wave packet}
\vspace*{-02mm}

In the experiments using effusive beams as a source of atoms, it is 
possible to describe the 
incident particles in terms of statistical mixtures of de Broglie 
waves having a well-defined 
momentum (plane waves). In that case, the reflection process at the 
evanescent wave mirror can be 
directly characterized using the results of the preceding sections. 
However, in some other 
situations (for example, when the incident atoms originate from an 
optical molasses where atom 
localization takes place [10]), it is more appropriate to describe 
the particles in terms of wave 
packets (superpositions of plane waves). In such a case, each partial 
plane wave of incident 
momentum $p_{\infty}$ experiences a different phase shift at 
reflection $\Delta \varphi_{\rm Schr}
(p_{\infty})$ (as given in Section 3.2), which shows up in a spatial 
shift of the center of the 
wave packet.

Let us consider an atomic wave packet incident on an evanescent wave 
mirror. In the asymptotic 
region, one may write the incident part of the atomic wave function 
$\psi_{\rm inc} (z,t)$ as
\begin{equation}
\psi_{\rm inc}(z,t) = \int dp_{\infty} \tilde{\psi}_{\rm inc} (p_{\infty}) 
\exp \left( - i \frac{p_{\infty}^2 t}
{2 M \hbar} - i \frac{p_{\infty} z}{\hbar} \right),
\end{equation}
\noindent
where $\tilde{\psi}_{\rm inc} (p_{\infty})$ denotes the Fourier transform 
of $\psi_{\rm inc}$. During the 
reflection process, each partial plane wave experiences a different 
phase shift $\Delta \varphi_{\rm Schr} (p_{\infty})$ [equation (19)], so that in the asymptotic 
region, the reflection part of 
the atomic wave function $\psi_{\rm ref} (z,t)$ reads
\begin{eqnarray}
&&\psi_{\rm ref} (z,t) = - \int dp_{\infty} 
\tilde{\psi}_{\rm inc}(p_{\infty}) \times 
\nonumber\\ 
&& {}\times 
\exp\left( - i \frac{p_{\infty}^2 
t}{2 M \hbar} + i \frac{p_{\infty} z}{\hbar} + i \Delta 
\varphi_{\rm Schr}(p_{\infty}) \right),
\end{eqnarray}
\noindent
where the minus sign in front of the integral results from our 
peculiar choise of phase origin.

By assuming that $\tilde{\psi}_{\rm inc}(p_{\infty})$ is peaked 
around 
the average momentum $\bar{p}_{\infty}$, it is possible to 
characterize the position $z_{\rm wp} (t)$ 
of the center of the wave 
packet (23) or (24) via the method of stationary phase. One readily 
finds
\begin{mathletters}
\begin{eqnarray}
z_{\rm wp}^{\rm inc} (t) &=& 
- \frac{\bar{p}_{\infty}}{M} t,
\\
z_{\rm wp}^{\rm ref} (t) &=& 
- \frac{\bar{p}_{\infty}}{M} t - \hbar \left( 
\frac{\partial \Delta \varphi_{\rm Schr} (p_{\infty})}{\partial 
p_{\infty}} \right)_{\bar{p}_{\infty}}, 
\end{eqnarray}
\noindent
where $z_{\rm wp}^{\rm inc} (z_{\rm wp}^{\rm ref})$ denotes the position of the 
center of the incident (reflected) 
wave packet. Equation (25) shows that, as far as the asymptotic wave 
packets are concerned, the 
evanescent wave mirror behaves as an infinitely steep effective 
mirror located at the position 
$\zeta_{\rm wp}$ given by 
\end{mathletters}
\begin{equation}
\zeta_{\rm wp} = -\frac{\hbar}{2} \left( \frac{\partial \Delta 
\varphi_{\rm Schr}}{\partial p_{\infty}} 
\right)_{\bar{p}_{\infty}} .
\end{equation}

Substituting for $\Delta \varphi_{\rm Schr}$ in (26) using (19) gives
\begin{equation}
\zeta_{\rm wp} = \kappa^{-1} (\ln (p_{\rm max} / 2 \hbar \kappa) - Re 
\Psi (1+i\bar{\alpha})),
\end{equation}
\noindent
where $\bar{\alpha} = \bar{p}_{\infty} / \hbar \kappa$ is the 
dimensionless parameter given by 
equation (16), and where $\Psi$ is the digamma function defined as
\begin{equation}
\Psi (x) = \partial \ln \Gamma (x) / \partial x.
\end{equation}

As in the preceding section, we distinguish between two limiting 
regimes of reflection of the 
atomic wave packet.

In the limit $\bar{\alpha} \gg 1$ (semiclassical regime of 
reflection, see Section 3.2), where 
equation (27) reduces to 
\begin{equation}
\bar{\alpha} \gg 1:  \zeta_{\rm wp} = \zeta_{\rm cl} (\bar{p}_{\infty}) + 
O(1/\bar{\alpha}^2),
\end{equation}
\noindent
the asymptotic wave packet appears to be instantaneously reflected at 
an effective mirror located 
in $z = \zeta_{\rm cl}$ [equation (5)]. The reflection process is 
thus analogous to that of a 
classical particle of incident momentum $\bar{p}_{\infty}$.%
\footnote{Within the framework of the 
WKB approach, it can be shown that this result holds for any mirror 
potential vanishing at large 
$z$.} Note that the position of the effective mirror for the atomic 
wave packet corresponds to 
$\zeta_{\rm cl}$, and not to $\zeta_{\rm WKB}$ [equation (10b)].

In the limit $\bar{\alpha} \ll 1$ (quantum regime of reflection), 
where equation (27) reads
\begin{equation}
\bar{\alpha} \ll 1:  \zeta_{\rm wp} \cong \zeta_{\rm Schr} (\bar{\alpha} \ll 
1) \cong \zeta_{\rm cl} (\hbar 
\kappa) + \gamma \kappa^{-1} ,
\end{equation}
\noindent
the evanescent wave mirror behaves as an infinitely steep potential 
barrier located at the 
position $z \approx \zeta_{\rm cl} (\hbar \kappa) + \gamma 
\kappa^{-1}$, which instantaneously 
reflects the atomic wave packet. The fact that this position 
coincides with that of the effective 
mirror for an atomic plane wave of incident momentum 
$\bar{p}_{\infty}$ [equation (22b)] is not 
surprising because in the quantum regime, the position of the 
effective mirror $\zeta_{\rm Schr}$ is 
independent of the incident atomic momentum (see Section 3.2). As a 
result, all partial plane 
waves of the wave packet are reflected at the same barrier, and the 
wave packet behaves in the 
same way.

The position $\zeta_{\rm wp}$ of the effective mirror describing the 
reflection process of the wave 
packet at the evanescent wave mirror is represented in Fig. 4 
together with $\zeta_{\rm cl}$ as a 
function of the dimensionless parameter $\bar{\alpha} = 
\bar{p}_{\infty} / \hbar \kappa$. One 
clearly distinguishes between the semiclassical and the quantum 
regime of reflection of the wave 
packet.

\begin{figure}[bth]
\centerline{\epsfig{file=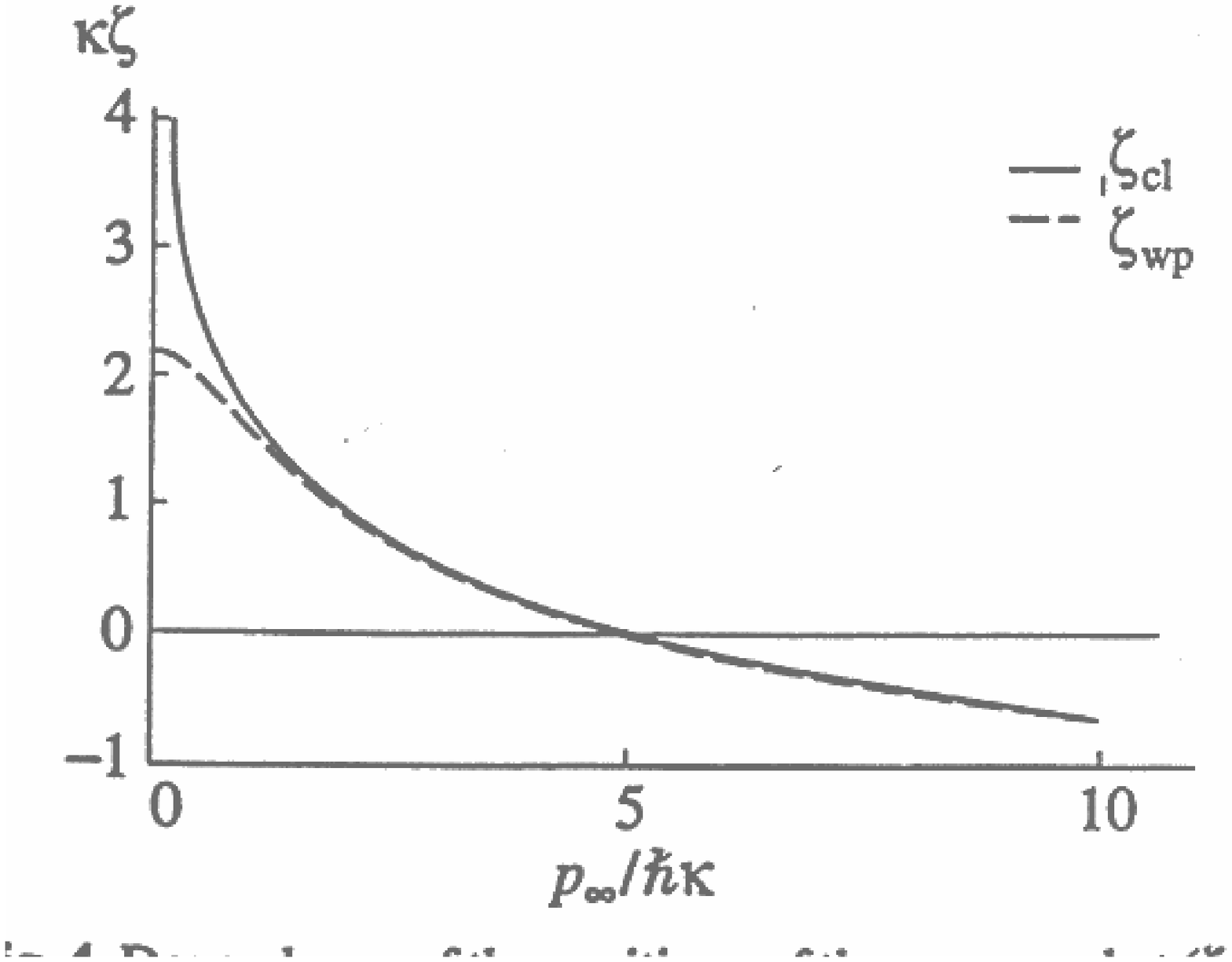,width=80mm,clip}}
    \caption[]{%
Dependence of the positions of the wave packet ($\zeta_{\rm wp}$) 
and the classical ($\zeta
_{\rm cl}$) effective mirrors vs.\ the dimensionless average incident 
momentum of the wave packet 
$\bar{\alpha} = \bar{p}_{\infty}/\hbar\kappa$ for $p_{\rm max} = 10 
\hbar \kappa$. $\zeta_{\rm cl}$ and $\zeta_{\rm wp}$ 
coincide in the semiclassical regime of reflection ($\bar{\alpha} \gg 
1$). By contrast, in the 
quantum regime of reflection ($\bar{\alpha} \ll 1$), $\zeta_{\rm wp}$ 
tends toward a constant, whereas 
$\zeta_{\rm cl}$ tends to infinity as $\bar{\alpha} \to 0^+$.
}
\end{figure}

\section{Conclusion}
\vspace*{-02mm}

We have presented a detailed theoretical investigation of the 
reflection process of an atomic de 
Broglie wave at an evanescent wave mirror in the regime of coherent 
atom optics. Our calculation 
of the atomic phase shift at reflection using the exact solution of 
the corresponding 
Schr\"odinger equation has allowed us to identify two limiting 
regimes of reflection. The 
semiclassical regime corresponds to incident de Broglie wavelengths 
much smaller than the decay 
length of the evanescent optical potential and can be satisfactorily 
accounted for by the WKB 
method. The evanescent wave mirror then behaves as a despersive 
dephasing mirror. In the quantum 
regime of reflection, where the incident atomic de Broglie wavelength 
is larger than the decay 
length of the evanescent potential, the evanescent wave mirror 
behaves as a nondispersive hard 
potential barrier located in front of the actual evanescent wave 
mirror surface.

In experiments using either a supersonic beam or laser-cooled atoms 
accelerated by the earth 
gravity field, the minimum achievable atomic incident momentum is 
typically of the order of 
$10 \hbar \kappa$. Under such conditions, the atomic reflection 
process can always be accounted 
for in semiclassical terms. However, the experimental observation of 
atomic reflection in the 
quantum regime seems feasible, using, for example, an evanescent wave 
mirror located at the summit 
of an atomic fountain where the atomic momentum approaches zero. 
Another, more challenging 
possibility would be to investigate the lowest bouncing modes of a 
gravitational cavity [6], which 
correspond to a de Broglie wavelength of the order of the decay 
length of the evanescent optical 
potential, and thus realize the quantum regime of reflection for the 
bouncing atoms.

The exact derivation of the atomic wave function presented in this 
paper [equation(17)] can serve 
as a starting point for the investigation of many other effects. 
These include tunneling through 
the optical potential barrier, the influence of atomic internal 
states on the reflection process, 
and, especially interesting, dipole-surface effects (such as the Van 
der Waals interaction), which 
may modify the potential we have assumed here, and hence also the 
phase shift at reflection.

\section*{Acknowledgements}
\vspace*{-02mm}

This work was supported by DRET (under Grant no. 91055) and the EEC 
(Science SC1-CT92-0778).

\appendix
\section{Experimental investigation of the reflection 
process}
\vspace*{-02mm}

We would like to comment in this appendix on the typical experimental 
parameters required to 
observe atomic reflection on an evanescent wave mirror. The regime 
of coherent atom optics is 
realized provided that the probability of spontaneous emission during 
reflection is negligible. It 
corresponds to the limit of small saturation of the atomic transition:
\begin{equation}
s = \frac{\Omega^2/2}{\Delta^2 + \Gamma^2/4} \ll 1,
\end{equation}
\noindent
where $\Omega = -d E_0 / \hbar$ is the resonant Rabi frequency that 
characterizes the coupling 
between the atomic dipole $d$ and the evanescent field of maximum 
amplitude $E_0$ (at the 
evanescent wave mirror surface), $\Gamma$ is the natural width of the 
atomic excited state, and 
$\Delta = \omega - \omega_A$ is the detuning between the frequency 
($\omega$) of the evanescent 
wave and the atomic ($\omega_A$) frequency. It also requires that the 
incident atom follows 
adiabatically the optical potential associated with the light-shifted 
ground-state level. This is 
achieved in the limit of large detuning from resonance (we assume 
$\Delta > 0$, which allows an atom
entering the optical potential in the ground-state to be reflected
at the evanescent wave mirror):
\begin{equation}
\Delta \gg \Gamma
\end{equation}
In the regime where (A1) and (A2) are fulfilled, it is possible to 
describe the atom reflection 
process by means of the Hamiltonian (1). Note, however, that for a 
given laser intensity, the 
maximum reflectible atomic momentum $p_{\rm max}$ decreases as the 
frequency detuning increases, as 
shown by the relation
\begin{equation}
\frac{p_{\rm max}^2}{2 M} = \frac{1}{2} \hbar \Delta s 
\end{equation}

In fact, by designing the evanescent wave mirror with the multilayer 
coating technique [12], 
it is possible fo fulfill conditions (A1) and (A2) while 
simultaneously being able to reflect 
atoms having high incident momentum ($p_{\rm max} \gg \hbar \kappa$). 
Typical experimental parameters 
are indicated in the table in the case of the D2 line of $^{85}$Rb.

\vskip 7.5mm

\noindent\small
Typical experimental parameters for atomic reflection at an 
evanescent wave 
mirror using the D2 line of $^{85}$Rb atoms
\begin{center}
\begin{tabular}{l|@{\hspace*{01mm}}c|@{\hspace*{02mm}}c} 
\hline
Physical parameters \rule[-3.5mm]{0pt}{0pt}
\rule{0pt}{5mm}
& 
\parbox[c]{07.5mm}{\centering
Nota-
\\
tion}
& 
\parbox[c]{15mm}{\centering
Typical
\\
value}
\\
\hline
\rule{0pt}{3.5mm}%
Laser wavelength & & $780\,{\rm nm}$ \\[0.5mm]
Natural linewidth & $\Gamma$ & $6\,{\rm MHz}$ \\[0.5mm]
Frequency detuning from resonance & $\Delta$ & $5 \times 10^4 
\,\Gamma$ \\[0.5mm]
Maximum intensity of the evanes- & & \\[-0.7mm]
cent wave & $E_0^2$ & $10^4 \,{\rm W}/{\rm cm}^2$ \\[0.5mm]
Saturation parameter & $s$ & $6 \times 10^{-4}$ \\[0.5mm]
Maximal reflected momentum & $p_{\rm max}$ & $150 \,\hbar \kappa$ 
\\[0.5mm]
Probability of spontaneous emission & & \\[-0.7mm]
per reflection for $p_{\infty} = p_{\rm max}$ & & $2.5 \times 
10^{-3}$ \\[0.5mm]
Probability of nonadiabatic depar- & & \\[-0.7mm]
ture from the light-shifted ground- & & \\[-0.7mm]
state level for $p_{\infty} = p_{\rm max}$ & & $\le 8 \times 
10^{-15}$ \\[0.5mm]
Decay length of the evanescent opti- & & \\[-0.7mm]
cal potential & $1/\kappa^{-1}$ & $\approx 100\,{\rm nm}$ \\[0.5mm]
\rule[-1.5mm]{0pt}{3.5mm}%
Reflection time for $p_{\infty} = p_{\rm max}$ & $\tau_{\rm refl}$ 
& $\approx 4 \,\Gamma^{-1}$ 
\\
\hline
\end{tabular}
\end{center}

\section{Solution of the stationary Schr\"odinger 
equation}
\vspace*{-02mm}

The general solution of the Bessel-type Schr\"odinger equation (15) 
is a linear combination of the 
Bessel functions $I_{\pm i\alpha} (u)$. These function are defined by 
[11]
\begin{equation} 
I_{\pm i\alpha} = \sum_{n=0}^{\infty} \frac{1}{n! \,
\Gamma (n + 1 \pm 
i\alpha)} \left( \frac{u}{2} \right)^{2n \pm i\alpha},
\end{equation}
\noindent
where $\Gamma$ denotes the Euler gamma function. They both diverge as 
$z \to - \infty \Leftrightarrow u \to + \infty$ according to [11]
\begin{equation}
u \to + \infty:  I_{\pm i \alpha} (u) \cong \frac{1}{\sqrt{2\pi u}} 
e^u (1 + O(1/u)). 
\end{equation}

As a result, the only linear combination of $I_{\pm i\alpha} (u)$ 
satisfying the boundary 
condition (i) of Section 3.1 corresponds to the difference of 
$I_{i\alpha} (u)$ and $I_{-i\alpha} 
(u)$. This difference is proportional to the Bessel-$K$ function:
\begin{equation}
K_{i\alpha} = \frac{\pi}{\sinh(\pi \alpha)} \frac{1}{2i} 
(I_{-i\alpha}(u) - I_{i\alpha}(u)),
\end{equation}
\noindent
whose asymptotic expansion is [11]:
\begin{equation}
u \to + \infty:  K_{i\alpha} (u) \cong \sqrt{\frac{\pi}{2u}} e^{-u} 
(1 + O(1/u)).
\end{equation}

Equation (B4) shows that the atomic wave function decays very rapidly 
(as $\exp [-(p_{\rm max} / \hbar 
\kappa) e^{-\kappa z}]$) inside the potential barrier.

We finally consider the boundary condition (ii) of the Section 3.1. 
In the asymptotic region 
$z \to + \infty \Leftrightarrow u \to 0^+$, the expansion of $I_{\pm 
i\alpha}(u)$ is given by the 
first term of the series expansion (B1):
\begin{eqnarray}
z \to + \infty:&& 
I_{\pm i\alpha} (u(z)) \cong 
\frac{1}{|\Gamma(1+i\alpha)|} 
\exp\left( \mp i p_{\infty} z / \hbar \right.
\nonumber\\
&& \quad \left.
\pm i \alpha \ln 
\left( \frac{p_{\rm max}}{2 \hbar \kappa} \right) \mp i \arg \Gamma 
(1+i\alpha) \right)
\nonumber\\
&&
\end{eqnarray}
\noindent
with [9]:
\begin{equation}
|\Gamma (1+i\alpha)| = \sqrt{\frac{\pi\alpha}{\sinh (\pi\alpha)}}. 
\end{equation}

Combining (B3), (B5), and (B6), one readly finds that the only 
solution of equation (15) satisfying the boundary conditions (i) and 
(ii) of Section 3.1 is:
\begin{equation} 
\psi_{\rm Schr}(z) = 
\sqrt{\frac{4M}{p_\infty}\frac{\pi\alpha}{\sinh(\pi\alpha)}} 
\frac{1}{2i} (I_{-i\alpha} (u(z)) - I_{i\alpha} (u(z))),
\end{equation}
\noindent
or equivalently
\begin{equation} 
\psi_{\rm Schr}(z) = \sqrt{\frac{4M}{\pi\hbar\kappa}\sinh(\pi\alpha)} 
K_{i\alpha} (u(z))
.
\end{equation}

\end{multicols}

\end{document}